\documentclass{PoS}
\usepackage{epsfig}
\usepackage{cite}
\newcommand{\postscript}[2]{\setlength{\epsfxsize}{#2\hsize}
   \centerline{\epsfbox{#1}}}
\newcommand{\comment}[1]{}

\usepackage[usenames,dvipsnames]{xcolor}
\definecolor{orange}{cmyk}{0,0.5,1,0}
\definecolor{rossoCP3}{cmyk}{0,.88,.77,.40}
\definecolor{graa}{rgb}{0.8,0.8,0.8}
\definecolor{blaa}{rgb}{0.2,0.2,0.6}

\title{Is there anybody out there?}

\ShortTitle{Is there anybody out there?}

\author{Luis A. Anchordoqui\\
Department of Physics \& Astronomy,  Lehman College, City University of
  New York, NY 10468\\
Department of Physics,
 Graduate Center, City University
  of New York,  NY 10016\\
Department of Astrophysics,
 American Museum of Natural History, NY
 10024\\
        E-mail: \email{luis.anchordoqui@gmail.com}}

\author{Susanna M. Weber\\
Mamaroneck High School, 1000 Boston Post Rd., Mamaroneck, NY 10543\\
E-mail: \email{susamawe@gmail.com}}

\author{\speaker{Jorge F. Soriano} \\
Department of Physics \& Astronomy,  Lehman College, City University of
  New York, NY 10468\\
Department of Physics,
 Graduate Center, City University
  of New York,  NY 10016\\  
        E-mail: \email{jfdezsoriano@gmail.com}}

      \abstract{ The Fermi paradox is the discrepancy between the strong likelihood of alien intelligent life emerging (under a wide variety of assumptions) and the absence of any visible evidence for such emergence. We use this intriguing unlikeness to derive an upper limit on the fraction of living intelligent species that develop communication technology $\langle \xi_{\rm biotec} \rangle$. $\langle \cdots \rangle$ indicates average over all the multiple manners civilizations can arise, grow, and develop such technology, starting at any time since the formation of our Galaxy in any location inside it. Following Drake, we factorize $\langle \xi_{\rm biotec} \rangle$ as the product of  the  fractions in which: {\it (i)}~life arises, {\it (ii)}~intelligence develops, and {\it (iii)}~communication technology is developed. This averaging procedure must be regarded as a crude approximation because the characteristics of the initial conditions in a planet and its surroundings may affect the three phenomena with high complexity. In this approximation, the number of communicating intelligent civilizations that exist in the Galaxy at any given time is found to be $N = \langle \zeta_{\rm astro} \rangle \, \langle \xi_{\rm biotec} \rangle  \,  L_\tau$, where  $\langle \zeta_{\rm astro} \rangle$ is the average production rate of potentially habitable rocky planets with a long-lasting ($\sim 4~{\rm Gyr}$) ecoshell and $L_{\tau}$ is the length of time that a typical civilization communicates. We estimate the production rate of exoplanets in the habitable zone and using recent determinations of the rate of gamma-ray bursts (GRBs) and their luminosity function, we calculate the probability that a life-threatening (lethal) GRB could make a planet inhospitable to life, yielding $\langle \zeta_{\rm astro} \rangle \sim 2 \times 10^{-3}~{\rm yr}^{-1}$.  Our current measurement of $N =0$ then implies $\langle \zeta_{\rm biotec} \rangle < 5 \times 10^{-3}$ at the 95\%C.L., where we have taken $L_\tau > 0.3~{\rm Myr}$ such that $c L_\tau \gg$ propagation distances of Galactic scales ($\sim 10~{\rm kpc}$), ensuring that any advanced civilization living in the Milky Way would be able to communicate with us.}

\FullConference{35th International Cosmic Ray Conference - ICRC217 -\\
		10-20 July, 2017\\
		Bexco, Busan, Korea}

\begin{document}

On May 25, 1961 President Kennedy's announcement to put a man on the
moon and bring him back safely before the end of the decade set the
advent of human exploration of space for NASA, culminating to the
landing on the Moon on July 16, 1969. It is difficult to believe that
this is the only time such an event has ever happened in the history
of the universe. On the other hand, if there is alien life capable of
pulling off such a feat we must ask, as Fermi did, {\it where is everybody?}~\cite{Jones}.

By adopting the starting point of a first approximation of the answer, we can write 
 the number of intelligent civilizations in our galaxy at any given time capable of releasing detectable signals of their existence into space using a quite simple functional form, 
\begin{equation}
 N = R_\star \ f_{\rm p}  \ n_{\rm e}  \ f_{\ell} \ f_{\rm i} \ f_{\rm c} \ L_\tau \,,
\label{Drake}
\end{equation}
where $R_\star$ is the average rate of star formation, $f_{\rm p}$ is the fraction of stars with planetary systems, $n_{\rm e}$ is the number of planets (per solar system) with a long-lasting ($\sim 4~{\rm Gyr}$)   ecoshell, $f_\ell$ is the fraction of suitable planets on which life actually appears, $f_{\rm i}$ is the fraction of living species that develop intelligence, $f_{\rm c}$ is the fraction of intelligent species with communications technology, and $L_\tau$ is the length of time such civilizations release detectable signals into space (i.e. the lifetime of the {\it communicative phase})~\cite{Drake}.

Following~\cite{Prantzos:2013hg} we  separate (\ref{Drake}) into its astrophysical and biotechnological factors
\begin{equation}
N = \langle \zeta_{\rm astro} \rangle \, \langle \xi_{\rm biotec} \rangle \, L_\tau \, ,
\end{equation}
where $\langle \zeta_{\rm astro} \rangle = R_\star \, f_{\rm p} \, n_{\rm e}$ represents the production rate of habitable planets with long-lasting ecoshell (determined through astrophysics) and $\langle \xi_{\rm biotec} \rangle = f_\ell \, f_{\rm i} \, f_{\rm c}$ represents the product of all chemical, biological and technological factors leading to the development of a technological civilization. $\langle \cdots \rangle$ indicates average over all the multiple manners civilizations can arise, grow, and develop such technology, starting at any time since the formation of our Galaxy in any location inside it. This averaging procedure must be regarded as a crude approximation because the characteristics of the initial conditions in a planet and its surroundings may affect $f_\ell$, $f_{\rm i}$, and $f_{\rm c}$  with high complexity.  In this work we estimate the production rate of exoplanets in the habitable zone and the rate of planetary catastrophes which  could threaten the evolution of life on the surface of these worlds. Armed with these estimates we use our current measurement of $N =0$ to set an upper limit on  $\langle \zeta_{\rm biotec} \rangle$.

The star formation rate in the Galaxy is estimated to be $\dot M_\star = 1.65 \pm 0.19~M_\odot~{\rm yr}^{-1}$~\cite{Licquia:2014rsa,Chomiuk:2011fc}. This estimate has been derived assuming the Kroupa initial mass function (IMF)~\cite{Kroupa:2002ky,Kroupa:2003jm}. The shape of this IMF is lognormal-like and exhibits a peak around $M/M_\odot \approx 0.4$~\cite{Lada}, suggesting there are roughly 2 stars per $M_\odot$. Altogether, this yields $R_\star \approx 3~{\rm yr}^{-1}$. Now, only 10\% of these stars are appropriate for harboring habitable planets. This is because the mass of the star $M_\star < 1.1 M_\odot$ to be sufficiently long-lived (with main sequence lifetimes larger than 4.5~Gyr) and $M_\star > 0.7 M_\odot$ to possess circumstellar habitable zones outside the tidally locked region~\cite{Seager}.\footnote{The habitable zone is the orbital range around a star within which surface liquid water could be sustained. Since water is essential for life as we know it, the search for biosignature gases naturally focuses on planets located in the habitable zone of their host stars. The habitable zone of the solar system looks like a ring around the Sun. Rocky planets with an orbit within this ring may have liquid water to support life. The habitable zone around a single star looks similar to the habitable zone in our Solar System. The only difference is the size of the ring. If the star is bigger than the Sun it has a wider zone, if the star is smaller it has a narrower zone.  It might seem that the bigger the star the better. However, the biggest stars have relatively short lifespans, so the life around them probably would not have enough time to evolve. The habitable zones of small stars face a different problem. Besides being narrow they are relatively close to the star. A hypothetical planet in such a region would be tidally locked. That means that one half of it would always face the star and be extremely hot, while the opposite side would always be facing away and freezing. Such conditions are not very favorable for life.} The frequency $\eta_\oplus$ of terrestrial planets in and the habitable zone of solar-type stars can be determined using data from the {\it Kepler} mission~\cite{Borucki:2010zz,Borucki:2011nn,Batalha:2012gh}. Current estimates suggest $0.15^{+0.13}_{-0.06} < \eta_\oplus < 0.61^{+0.07}_{-0.15}$~\cite{Dressing:2013mid,Kopparapu:2013xpa}. The production rate of habitable planets is then $\sim 3~{\rm yr}^{-1} \times 0.1 \times 0.15 = 0.045~{\rm yr}^{-1}$.

Next, in line with our stated plan, we estimate a rough probability that a habitable planet will survive and remain in a habitable zone to present day. Gamma-ray bursts (GRBs) are short-lived, luminous explosions, thought to originate from relativistic plasma launched at the deaths of massive stars. The widely accepted interpretation of GRB phenomenology is that the observable effects are due to the dissipation of the kinetic energy of a relativistically expanding fireball~\cite{Piran:1999kx}. The physical conditions in the dissipation region produce a heavy flux of photons with energies above about 100~keV. It has been suggested that a nearby (Galactic) GRB may destroy the ozone layer, possibly making it damaging to life on Earth.\footnote{Ozone (O$_3$) forms a kind of layer in the stratosphere, which normally prevents about 90\% of the solar UVB ($280 - 315~{\rm nm}$) radiation from reaching the Earth's surface. UVB is extremely damaging to most organisms, particularly since it easily damages DNA and proteins.} Because of this, GRBs have been proposed to explain events of massive life extinction~\cite{Ruderman,Thorsett:1995us,Dar:1997he,Thomas:2004vj,Thomas:2005gpa,Melott:2003rs,Piran:2014wfa}.

GRBs are generally divided in two groups according to their duration: long $( > 2~{\rm s}$) and short ($< 2~{\rm s}$). Short GRBs are weaker and hence  their life threatening effect is negligible~\cite{Piran:2014wfa}. Throughout we consider only long GRBs.
We want to estimate the expected number of GRBs that can terminate life in a planet that is situated at a Galactic radius $R$. To this end we should take into account the following considerations:
\begin{description} 

\item [The luminosity-rate function]$\phi(L)$, which measures the number of GRBs with a luminosity in a small range around a given value $L$ occurring per unit time and volume, is given by
\begin{equation}
\phi(L) = n \left\{ \begin{array}{ll}
(L/L^*)^{-\alpha} & ~~~~~~~~~L_{\rm min} < L < L^* \\
(L/L^*)^{-\beta} & ~~~~~~~~~L^* < L < L_{\rm max}  \\
\end{array}
\right. ,
\label{eq:lumrate}
\end{equation}
where $\alpha = 1.2^{+0.2}_{-0.1}$, $\beta = 2.4^{+0.3}_{-0.6}$, $L^* = 10^{52.5\pm0.2}~{\rm erg \, s}^{-1}$, $L_{\rm min} = 10^{49}~{\rm erg \, s^{-1}}$, and $L_{\rm max} = 10^{54}~{\rm erg \, s^{-1}}$~\cite{Wanderman:2009es}. Here, $n$ is the volumetric rate of GRBs at $L=L^*$. We consider a fiducial value of $n_0 = 0.15^{+0.7}_{-0.8}\,\mathrm{yr^{-1}\,Gpc^{-3}}$~\cite{Wanderman:2009es}. To accommodate the metallicity bias determined in~\cite{Jimenez:2013dka} we follow~\cite{Piran:2014wfa} and take correction of a factor $10$.  It has been noted in~\cite{Gowanlock:2016mkk} that such a low metallicity correction factor yields an upper limit on the volumetric rate of long GRBs $n\leq0.1\,n_0$. To derive our upper bound on $\langle \xi_{\rm biotec} \rangle$ we will adopt $n =0.1 n_0$, since the larger the number of GRBs, the smaller the planets with long-lasting ecoshell, and therefore the larger the value of $\langle \xi_{\rm biotec} \rangle$. In general, any GRB regardless of its luminosity could terminate life if it is close enough to a planet. This is taken into account in our next point.

\item [The fluence] $\mathcal F$  measures the amount of energy per unit area arriving to a planet from a distant GRB.  If the distance between the GRB and the planet is $r$, an isotropic emission and conservation of energy implies \begin{equation} \mathcal{F}=\frac{E}{4\pi r^2}, 
\label{cuatro}
\end{equation} where $E$ is the total energy released by the GRB.

The effects that a copious flux of gamma rays may cause on the  Earth atmosphere have been  studied in~\cite{Thomas:2004vj,Thomas:2005gpa}. A fluence of $10~{\rm kJ/m}^2$ could cause a depletion of roughly 68\% of the ozone layer on a time scale of a month, whereas fluences of $100~{\rm kJ/m}^2$ and $1000~{\rm kJ/m}^2$ could cause depletions of about 91\% and 98\%, respectively. This implies that  a fluence of $10~{\rm kJ/m}^2$ could cause some damage to life, while $1000~{\rm kJ/m}^2$ will wipe out nearly the whole atmosphere causing a catastrophic life extinction event. Note, however, that the complete removal of {\it all life} on an Earth-like planet is a very unlikely event~\cite{Sloan}. The critical fluence $\mathcal F_c$  gives the limit on acceptable fluence on a planet. Following~\cite{Piran:2014wfa},  in our calculations we take $\mathcal F_c = 100~{\rm kJ/m}^2$ as our fiducial life threatening critical fluence.

\item [The hazard distance] of a GRB, characterized by its total energy $E$, measures the length for which the fluence is higher than the critical fluence $\mathcal F_c$.  Any planet within this distance will have life terminated.  For a fixed GRB energy and critical fluence, we use (\ref{cuatro}) to obtain the hazard distance \begin{equation} d(L,\mathcal F_c)=\sqrt{\frac{E}{4\pi\mathcal F_c}} = \sqrt{\frac{L\, \tau}{8\pi \mathcal F_c}} \,,\end{equation} where in the second rendition we have assumed that the total GRB energy is roughly the average ($\sim$ half) of the peak flux $E= L \, \tau /2$~\cite{Piran:2014wfa}. 
A good but rough estimate of $E$ follows from the assumption that the typical duration of long GRBs is $\tau \sim 20~{\rm s}$.

\begin{figure}[tbp] \begin{minipage}[t]{0.49\textwidth} \postscript{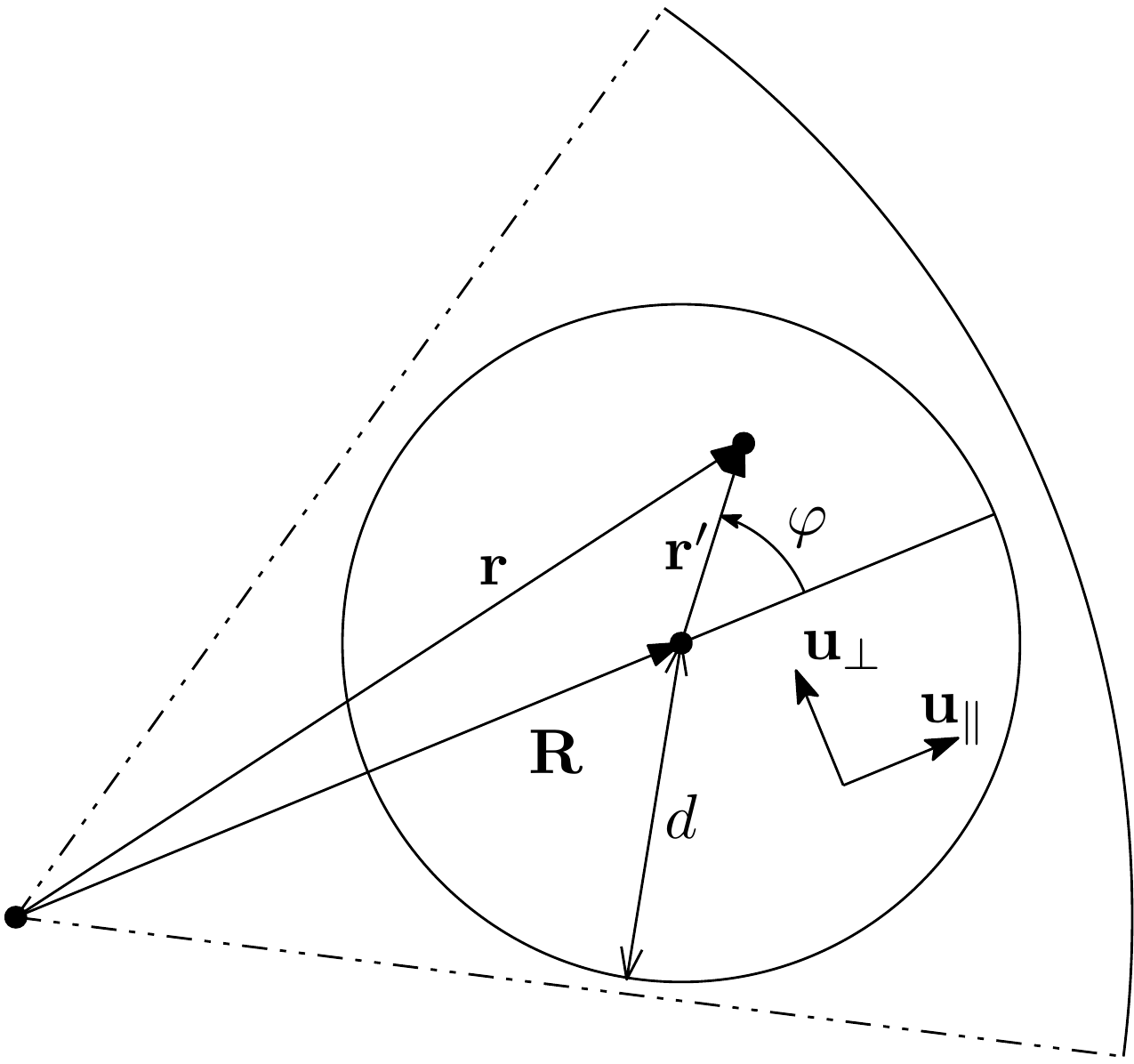}{0.7} \end{minipage} \hfill \begin{minipage}[t]{0.49\textwidth} \postscript{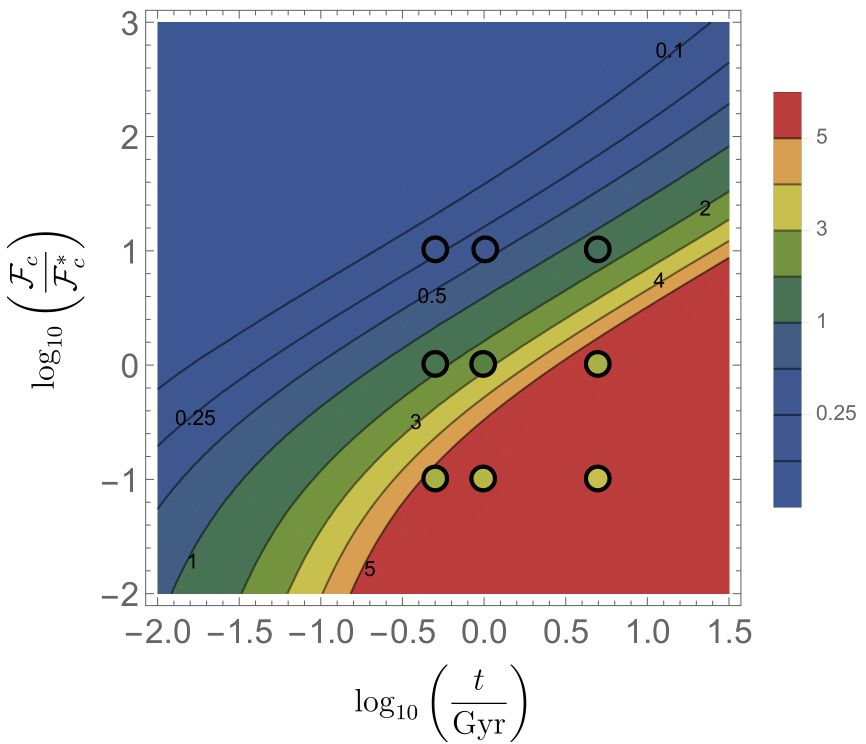}{0.99} \end{minipage} \caption{Hazard region (left) and probability contours for life destructing GRB on Earth as function of total time $t$ and critical fluency $\mathcal F_c$ normalized to $\mathcal F_c^*=100\,\mathrm{kJ/m^2}$ (right). We have taken $R_{\rm solar\, system} \approx 8.12~{\rm kpc}$. \label{fig:1}}
\end{figure}

\item [The fraction of hazardous galaxy] $p[d,R]$ measures the fraction of the total galactic mass that is within the hazard distance $d$,  for any point at radius $R$ from the Galactic center. The fraction of galactic mass that is contained within a surface element at radius $r$ is given by,
\begin{equation}
\rho(r)= \frac{1}{2\pi r_d^2} \,e^{-r/r_d} \,,
\label{massP}
\end{equation}
such that $\int \rho(r)\,\mathrm da=1$, with $r_d = 2.15 \pm 0.14~{\rm kpc}$~\cite{Bovy:2013raa}. As illustrated in Fig.~\ref{fig:1},
to calculate the fraction of hazardous galaxy around a point with radius $R$, one has to integrate (\ref{massP}) in a circular region of radius $d$ centred at $\mathbf R$,
\begin{equation}
p[d,R]= \frac{1}{2\pi r_d^2} \ \int_{S} e^{-r/r_d} \ \mathrm da \,.
\end{equation}
Defining the unit vectors $\mathbf u_\parallel=\mathbf R/R$ and $\mathbf u_\perp$ perpendicular to $\mathbf u_\parallel$, 
we can write $\mathbf R=R\,\mathbf u_\parallel$ and $\mathbf r'=r'(\cos\varphi\,\mathbf u_\parallel+\sin\varphi\,\mathbf u_\perp)$. Since $\mathbf  r=\mathbf R+\mathbf r'$,
\begin{equation}
r=\left|\mathbf r\right|=\sqrt{R^2+r'^2+2r'R\cos\varphi} \, .
\end{equation}
Defining $r'\equiv z\,d$, we have $\mathrm da=d^2\,z\,\mathrm dz\,\mathrm d\varphi$, where $z$ runs from $0$ to $1$, so that 
\begin{equation}
p[d,R]=\frac{d^2}{2\pi r_d^2}\int_0^{2\pi}\mathrm d\varphi\int_0^1\mathrm d z\,z\ \exp\left(-\frac{1}{r_d}\sqrt{R^2+d^2z^2+2d\,Rz\,\cos\varphi} \right).
\end{equation}
\item [The rate of life-threatening GRBs] at any position of the galaxy, specified by the radius $R$, is given by 
\begin{equation}
\Gamma(R,\mathcal F_c)=\frac{V(M_\star) }{L^*}\int_{L_{\rm min}}^{L_{\rm max}}\phi(L)  \ p[d(L,\mathcal F_c),R]\ {\rm d}L \,.
\label{doce}
\end{equation} 
The cosmological volume of a galaxy, with stellar mass $M_\star$, is defined by $V(M_\star) = M_\star/\rho_\star (z)$, where $\rho_\star (z) =10^{17.46 - 0.39 z} M_\odot~{\rm Gpc}^{-3}$ is the average stellar density as a function of redshift $z$~\cite{Li:2015dsa}.
For a galaxy like our own Milky Way, $M_\star \approx 6 \times 10^{10} M_\odot$~\cite{McMillan:2011wd} and so at $z=0$ we have $V(M_\star) \sim 10^{-7}~{\rm Gpc}^3$.
\end{description}

Taking into account all of these considerations we now turn to estimate  the expected number of GRBs that can terminate life in a planet that is situated at a Galactic radius $R$. As it turns out $\Gamma (R,\mathcal F_c)$ is ${\cal O} (\mathrm{Gyr^{-1}})$. This means that  for $\mathrm{Gyr}$ time scales we expect to have (on average) a small number of  GRB events. This kind of estimate is then well suited to Poisson statistics. The Poisson distribution is a discrete probability distribution for the counts of events that occur randomly in a given interval of time (or space). Of particular interest here, the probability of having $i$ GRBs  when the expected average is $\mu$ is given by  $p_i=e^{-\mu} \mu^i/i!$. The probability of having $1$ or more GRBs is $p\equiv \sum_{i=1}^\infty p_i =1-p_0 =1-e^{-\mu}$. The average number of GRBs during a time $t$ is  $\mu=\Gamma(R,\mathcal F_c) \  t$, and therefore \begin{equation} p(t,R,\mathcal F_c)=1-e^{-\Gamma(R,\mathcal F_c) \  t}.
\label{trece}
\end{equation}

In Fig.~\ref{fig:1} we show probability contours of at least one GRB having occurred in the past time $t$ with enough flux to produce significant life extinction on Earth.  Since the probabilities get too high in most of the parameter space of interest, the legend does not show the probability $p$ (which would be too close to $1$), but instead the parameter $k$ defined by 
\begin{equation}
p=\frac{1}{\sqrt{2\pi}}\int_{-k}^{+k} e^{-x^2/2} \, {\rm d} x \,  .
\label{kp}
\end{equation}
(\ref{kp}) gives the usual correspondence between probabilities and the standard deviation for a normal distribution, which leads to $k=\sqrt{2} \ \mathrm{Erf}^{-1}(p)$. Some value correspondences between $k$ and the probabilities in $\%$ are $k=1 \leftrightharpoons 68\%$, $k=2\leftrightharpoons 95.5\%$, and $k=5\leftrightharpoons 99.99994\%$.  The circles indicate selected values of ${\cal F}_c/({\rm kJ/m^2}) = 10, 100, 1000$ and $t/{\rm Gyr} = 0.5, 1, 5$. Our estimates are in good agreement with those given in Table~II of~\cite{Piran:2014wfa}.  These findings seem to indicate that a nearby GRB may have caused one of the five greatest mass extinctions on Earth.  However, it is important to remind the reader that the estimates shown in Fig.~\ref{fig:1} rely on an upper limit of the volumetric rate of long GRBs; namely, $n = 0.1 \, n_0$.

\begin{figure}[tbp]
\begin{minipage}[t]{0.49\textwidth}
\postscript{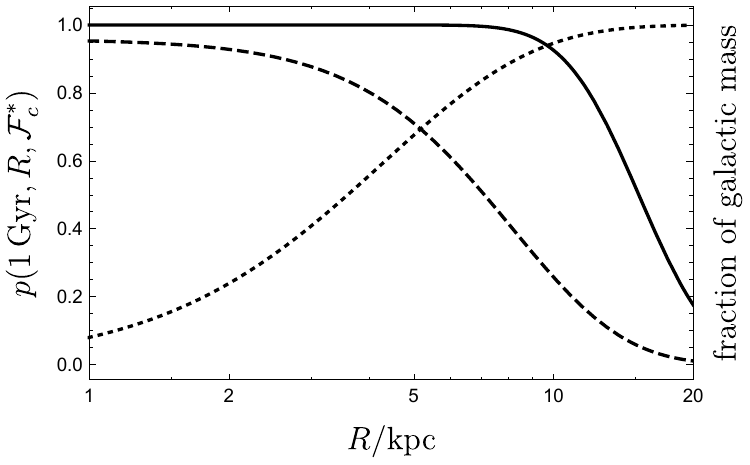}{0.99}
\end{minipage}
\hfill
\begin{minipage}[t]{0.49\textwidth}
\postscript{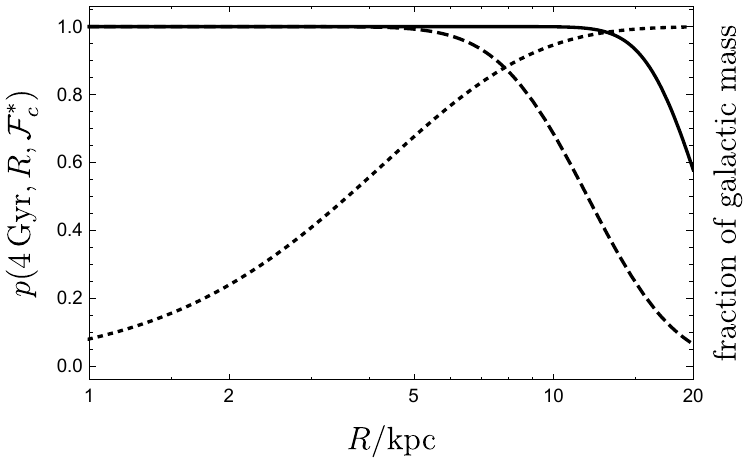}{0.99}
\end{minipage}
\caption{
Probability of having one or more lethal GRBs in $1\,\mathrm{Gyr}$ (left) and $4~{\rm Gyr}$ (right) for a critical fluency $\mathcal F_c= 100~\mathrm{kJ/m^2}$, as a function of distance to the Galactic center. The  solid line  corresponds to $n = n_0$ and the long-dashed to $n = 0.1 n_0$. Following the right legend, the short-dashed line measures the total amount of mass enclosed in a radius lesser than $R$.}\label{fig:2}\end{figure}

Coming back to our calculation, we can use (\ref{trece}) to compute the probability of having at least one lethal GRB, for a critical fluency ${\cal F}_c$ and a fixed lookback time $t_c$, as a function of the distance $R$. Life has been evolving on Earth for close to 4~Gyr~\cite{Mojzsis,Dodd}, but complex life is well under 1~Gyr old, and intelligent life is only a Myr old at most. In what follows we adopt $t_c =1~{\rm Gyr}$ and 4~Gyr as critical time intervals for life evolution~\cite{Loeb:2016vps}. In Fig.~\ref{fig:2} we show the probability of having one or more lethal GRBs  for a critical fluency $\mathcal F_c= 100~\mathrm{kJ/m^2}$ and $t_c/{\rm Gyr} = 1$ and 4, as a function of distance to the Galactic center.

\begin{figure}[tbp]
\begin{minipage}[t]{0.49\textwidth}
\postscript{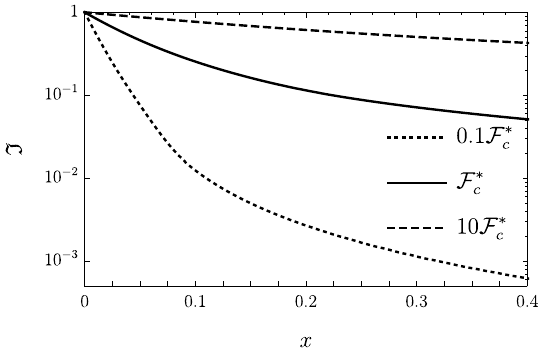}{0.99}
\end{minipage}
\hfill
\begin{minipage}[t]{0.49\textwidth}
\postscript{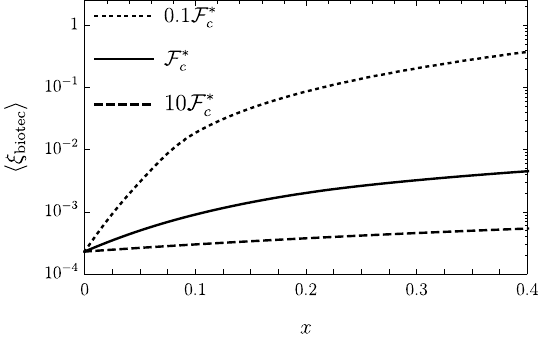}{0.99}
\end{minipage}
\caption{The integral $\mathfrak I$ (left) and the variation of the upper limit on $\langle \xi_{\rm biotec} \rangle$ (right) as a function of $x$, for three different hazardous fluences. \label{fig:3}}
\end{figure}

All we need to do now is add the components together to arrive at the production rate of habitable planets with a long-lasting ecoshell,
\begin{equation}
\langle \zeta_{\rm astro} \rangle = 0.045~{\rm yr}^{-1} \int_0^\infty \int_0^{2\pi}  \left[1 - p (t,R,{\cal F}_c) \right] \ \rho(R)  \ {\rm d}\varphi \ R \, {\rm d}R \, .
\label{quince}\end{equation}
It is of interest to study how $\langle \zeta_{\rm astro} \rangle$ depends on the different parameters involved. According to (\ref{eq:lumrate}), (\ref{doce}) and (\ref{trece}), the dependence of $p$ in $n$ and $t$ can be grouped in the same functional dependence. Introducing $\left. \Gamma_0 \equiv \Gamma \right|_{n=n_0}$, one can rewrite $\Gamma\,t=x\,\Gamma_0\,t_0$, where $x=(t/t_0)(n/n_0)$  and $t_0=1\,\mathrm{Gyr}$. We can then rewrite (\ref{quince})  as $\langle\zeta_{\mathrm{astro}}\rangle=0.045\,\mathrm{yr^{-1}} \mathfrak I(x)$, where
\begin{equation}
\mathfrak I(x)=2\pi\int_0^\infty \exp[-x\,\Gamma_0(R,\mathcal F_c)\,t_0] \ \rho(R) \ R \, {\rm d}R \, .
\label{eq:}\end{equation}
The value of $\mathfrak I(x)$ is shown in Fig.~\ref{fig:3} as a function of $x<0.4$ for different values of $\mathcal F_c$. Note that for $\mathcal F_c= 100~\mathrm{kJ/m^2}$, with $t_c = 1~{\rm Gyr}$ and $4~{\rm Gyr}$, (\ref{quince})  leads to $\langle \zeta_{\rm astro} \rangle = 1 \times 10^{-2}~{\rm yr}^{-1}$ and $2\times 10^{-3}~{\rm yr}^{-1}$, respectively.

Finally, to determine the upper bound on $\langle \xi_{\rm biotec} \rangle$ we must decide on the possible minimum $L_\tau$. Herein we consider $L_\tau > 0.3~{\rm Myr}$ such that $c L_\tau \gg$ propagation distances of Galactic scales ($\sim 10~{\rm kpc}$). This would provide enough time to receive electromagnetic (and/or high-energy neutrino~\cite{Learned:2008gr}) signals from any advanced civilization living in the Milky Way which is trying to communicate with us.

As of today, the non-observation of evidence of  advanced civilizations implies that models of $\langle \xi_{\rm biotec} \rangle$ predicting $N > 3.09$ are excluded at the 95\%~C.L.~\cite{Feldman:1997qc}.\footnote{If a corresponding hypothesis test is performed, the confidence level (C.L.) is the complement of the level of statistical significance, e.g, a 95\% confidence interval reflects a significance level of 0.05.}  Assuming that evolution requires 4~Gyr for life to evolve and that the communication phase with advanced civilizations must last at least 0.3~Myr, we obtain $\langle \xi_{\rm biotec} \rangle < 5 \times 10^{-3}$ at the 95\%~C.L. If instead we consider that only 1~{\rm Gyr} would be required (on average) for intelligent life to evolve the 95\% C.L. upper limit becomes more restrictive:  $\langle \xi_{\rm biotec} \rangle < 1 \times 10^{-3}$.  The dependence of $\langle \xi_{\rm biotec} \rangle$ with $x$ is shown in Fig.~\ref{fig:3}. A closing argument is that our estimate for the production rate of habitable planets is overly conservative, as we have adopted the present-day star formation rate. It has been noted that the average star formation rate in the Galaxy could be about 4 times the current rate~\cite{Kennicutt:2012ea}. 

In summary, in this paper we have derived an upper bound on the average fraction of living intelligent species that develop communication technology: $\langle \xi_{\rm biotec} \rangle < 5 \times 10^{-3}$ at the 95\%~C.L. Future observations 
could help to tighten this bound. In particular,  a new arsenal of data will certainly provide an ideal testing ground to improve our understanding about: {\it (i)}~the occurrence of exoplanets in the habitable zone, {\it (ii)}~the early star formation rate models, and {\it (iii)}~the GRB phenomenology. The past few years have witnessed  the discovery of more and more rocky planets that are larger and heftier than Earth. Finding the  {\it Earth-twins}  is a higher order challenge, because these smaller planets produce fainter signals and hence only a few have been discovered. Technology to detect and image Earth-like planets has been developed  for use of  the next generation space telescopes.  The Transiting Exoplanet Survey Satellite (TESS) is NASA's next step in the search for planets outside of our solar system, including those that could support life. The NASA roadmap will subsequently continue  with the launch of the  James Webb Space Telescope (JWST)~\cite{jwst} and perhaps the proposed Wide Field Infrared Survey Telescope - Astrophysics Focused Telescope Assets (WFIRST-AFTA) early in the next decade~\cite{wfirst}.  The ability to detect alien life may still be years or more away, but the quest is underway. \\

We thank Michael Unger for discussion and Pink Floyd for inspiration.
This research was supported by the U.S. NSF (Grant PHY-1620661) and NASA (Grant  NNX13AH52G).

%\section{...}

\end{document}